\documentclass[journal]{vgtc}                
\ifpdf
  \pdfoutput=1\relax                   
  \usepackage{graphicx}                
  \DeclareGraphicsExtensions{.pdf,.png,.jpg,.jpeg} 
\else
  \usepackage{graphicx}                
  \DeclareGraphicsExtensions{.eps}     
\fi%

\graphicspath{{figures/}{pictures/}{images/}{./}} 

\usepackage{microtype}                 
\PassOptionsToPackage{warn}{textcomp}  
\usepackage{textcomp}                  
\usepackage{mathptmx}                  
\usepackage{times}                     
\usepackage{cite}                      
\usepackage{tabu}                      
\usepackage{booktabs}                  

\usepackage{url}

\preprinttext{To appear in IEEE Transactions on Visualization and Computer Graphics.}

\onlineid{0}

\vgtccategory{Research}
\vgtcpapertype{please specify}

\title{Is Projection Mapping Natural?\\
\LARGE{Towards Physical World Augmentation Consistent with Light Field Context}}

\author{Daisuke Iwai}
\authorfooter{
\item
 Daisuke Iwai is with Graduate School of Engineering Science, Osaka University.
}

\shortauthortitle{Iwai: Is Projection Mapping Natural?}

\abstract{Projection mapping seamlessly merges real and virtual worlds. Although much effort was made to improve its image qualities so far, projection mapping is still unnatural. We introduce the first steps towards natural projection mapping by making the projection results consistent with the light field context of our daily scene.%
} 


\CCScatlist{ 
 \CCScat{K.6.1}{Management of Computing and Information Systems}%
{Project and People Management}{Life Cycle};
 \CCScat{K.7.m}{The Computing Profession}{Miscellaneous}{Ethics}
}

\vgtcinsertpkg

\begin{document}


\firstsection{Introduction}

\maketitle

Projection mapping is a spatial augmented reality (AR) approach that manipulates the appearance of a physical object by projected imagery as if its surface reflectance property dynamically changes~\cite{10.5555/1088894,doi:10.1111/cgf.13387}. Researchers have considered that this illusionary effect is useful in various application fields including but not limited to teleconferencing~\cite{10.1145/280814.280861,10.1145/2818048.2819965,8172039}, museum guide~\cite{1377099,SCHMIDT20191}, makeup~\cite{doi:10.1111/cgf.13128,8007312}, object search~\cite{Iwai2011,8007248,10.1145/1959826.1959828,10.1145/1015706.1015738}, product and architecture design~\cite{CASCINI2020103308,6949562,10.1111:j.1467-8659.2011.02066.x,8797923,1544657}, urban planning~\cite{10.1145/302979.303114}, artwork creation~\cite{970539,10.1145/1166253.1166290,10.1145/2366145.2366176}, medicine~\cite{00000658-201806000-00024,kijima07}, and entertainment~\cite{6193074}.

An advantage of the projection mapping compared to the other types of AR displays (e.g., optical and video see-through head-mounted displays) is that users can observe the augmentations on neighboring physical surfaces without holding and wearing any devices. On the other hand, the surfaces can be arbitrary shapes and have spatially varying reflectance properties, and thus, the image quality of the projected image is easily degraded. For example, a non-planar surface deforms projected imagery, and a textured surface distorts the projected colors. Researchers have tackled these problems and developed various image correction techniques that successfully solved the issues [1,2]. However, even though the image correction techniques are applied, the appearance of the projected object is still not visually satisfactory in terms of naturalness.

\section{Primary Factors Causing Unnaturalness}

\begin{figure}[t]
  \centering
  \includegraphics[width=0.98\hsize]{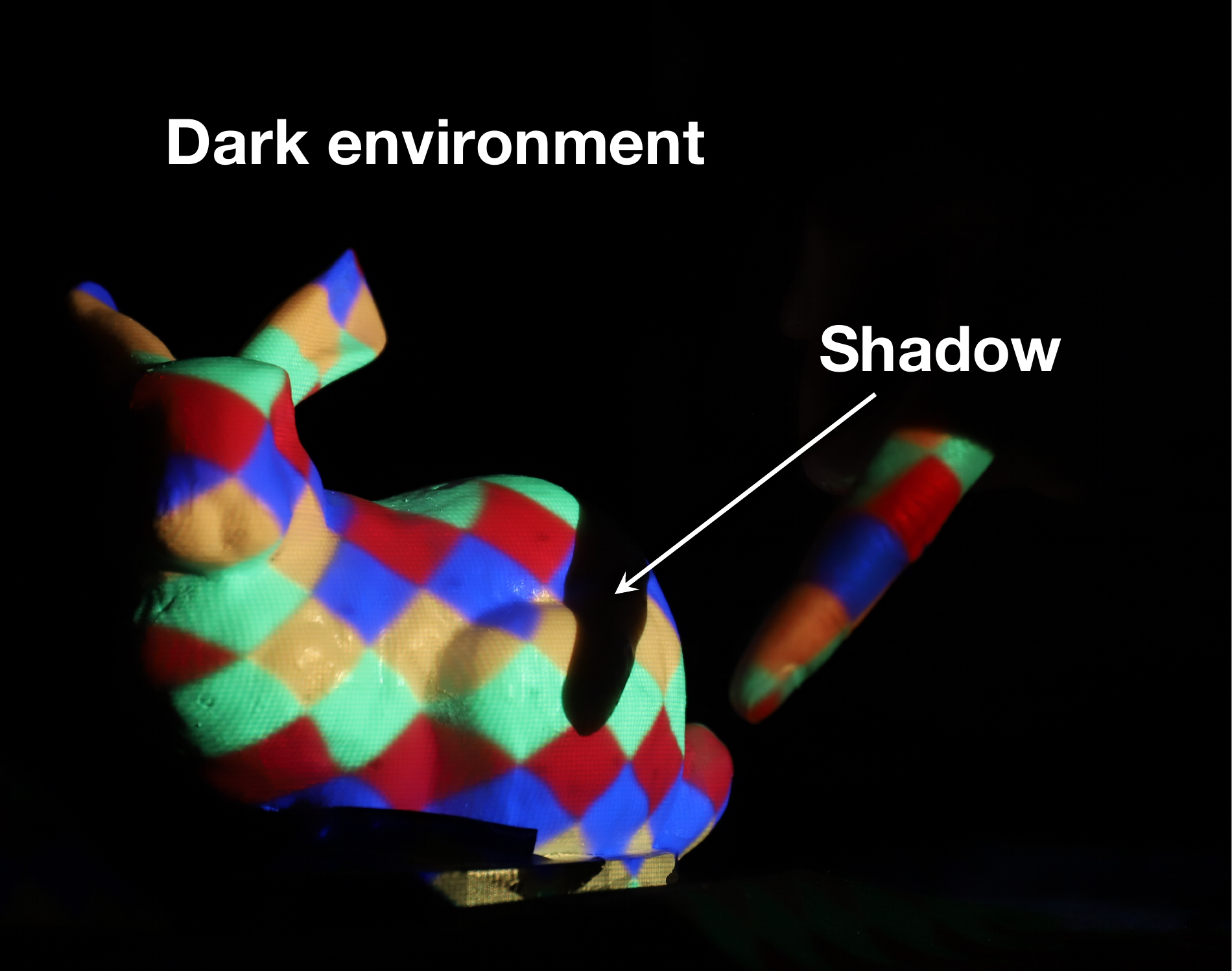}
  \caption{Primary factors causing unnaturalness in projection mapping.}
  \label{fig1}
\end{figure}

\autoref{fig1} shows the two primary factors causing unnaturalness in projection mapping. First, projection mapping works only in a dark environment. If there is an environment light, the contrast of the projected result significantly decreased, and essential image details would be diminished. Therefore, we always experience projection mapping in a dark environment where only a projected object is brightly lit. Human observers perceive that the object emits light rather than that the surface reflectance property is changed. Second, if a user locates between the projection object and the projector, they occlude the projected light, which casts its shadow on the object where no graphical information is visible. This also hinders the illusory effect of the projection mapping as if the surface reflectance property changes.

The two factors significantly degrade the naturalness of the projection mapping results because these are not consistent with the light field context in our daily environment. We usually observe physical objects under environmental lighting, where the scene around the objects is not dark. When we approach the objects, they are not completely occluded from the environment light sources, and thus, the surface textures are still visible. These light field contexts should be maintained in projection mapping to provide natural AR experiences. In this invited talk, I will introduce a technical solution to overcome each of these factors.

\section{Technical Solutions}

\begin{figure}[t]
  \centering
  \includegraphics[width=0.98\hsize]{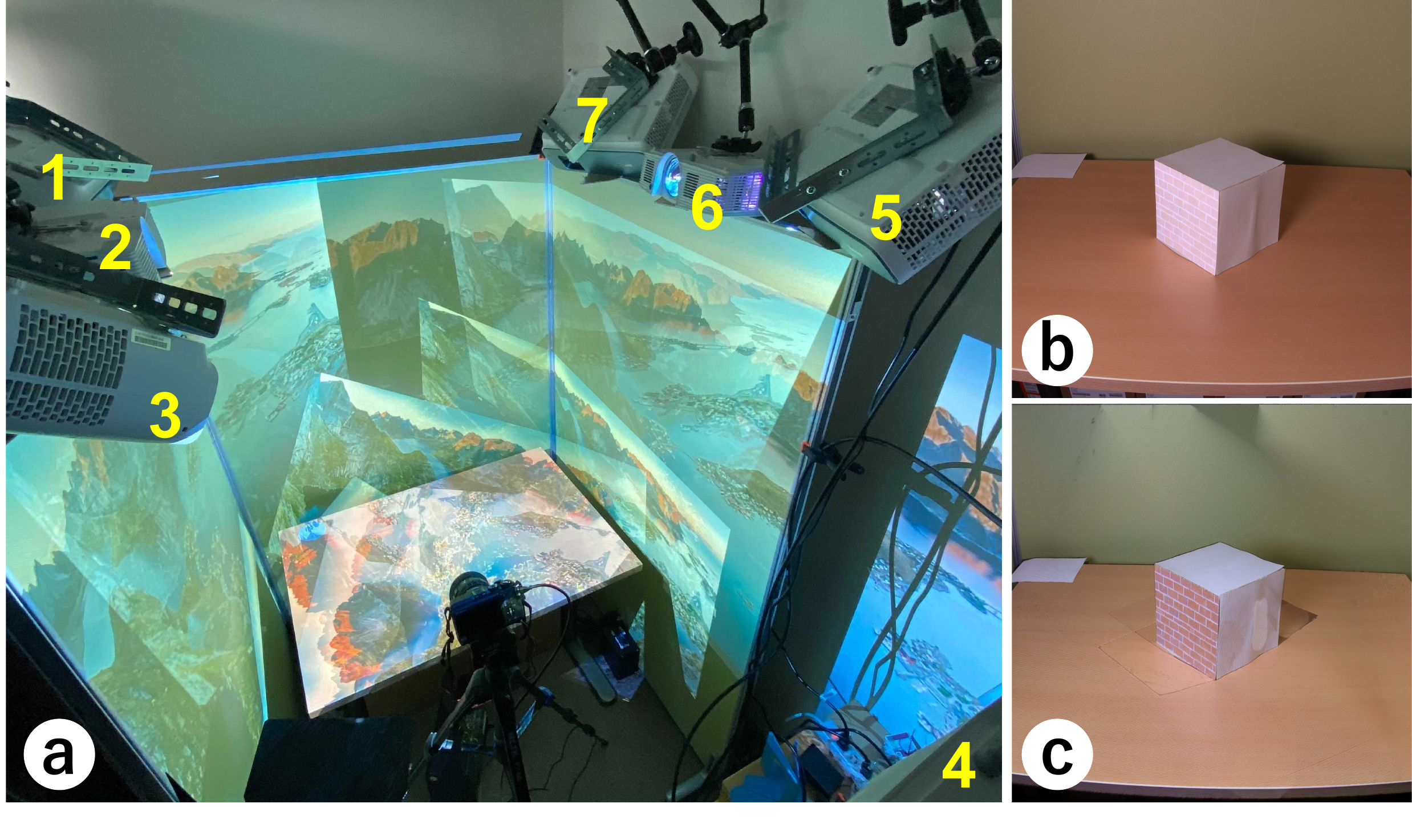}
  \caption{Multi-projection system for reproducing room lights. (a) Experimental setup. (b) Projection mapping result (brick pattern projection onto a white cube) under room lights. (c) Projection mapping result using our proposed system reproducing the light field of the room lights.}
  \label{fig2}
\end{figure}

A technical solution for the dark environment problem is replacing room lights with projectors and reproducing the light field produced by the room lights using the projectors. Multiple projectors are required to illuminate the entire room's surfaces. In a pilot study, we prepared a small booth and installed seven projectors (\autoref{fig2}a). We captured the surface appearance under room lights using a 360-degrees camera. Then we turned off the room lights and reproduced the surface appearance using the projectors. \autoref{fig2}b shows a projection mapping result under the room lights, where a white cube was the target object on which a brick texture is projected. \autoref{fig2}c shows a result in a condition where our projector-based room light reproduction was applied. We found that the brick texture was clearer in \autoref{fig2}c than in \autoref{fig2}b. The contrast of the projected result was significantly degraded by the room lights in \autoref{fig2}b, while that was not degraded in \autoref{fig2}c. We also confirmed that the appearance of the other surfaces than the cube was also well reproduced by the projected imagery in \autoref{fig2}c.

\begin{figure}[t]
  \centering
  \includegraphics[width=0.98\hsize]{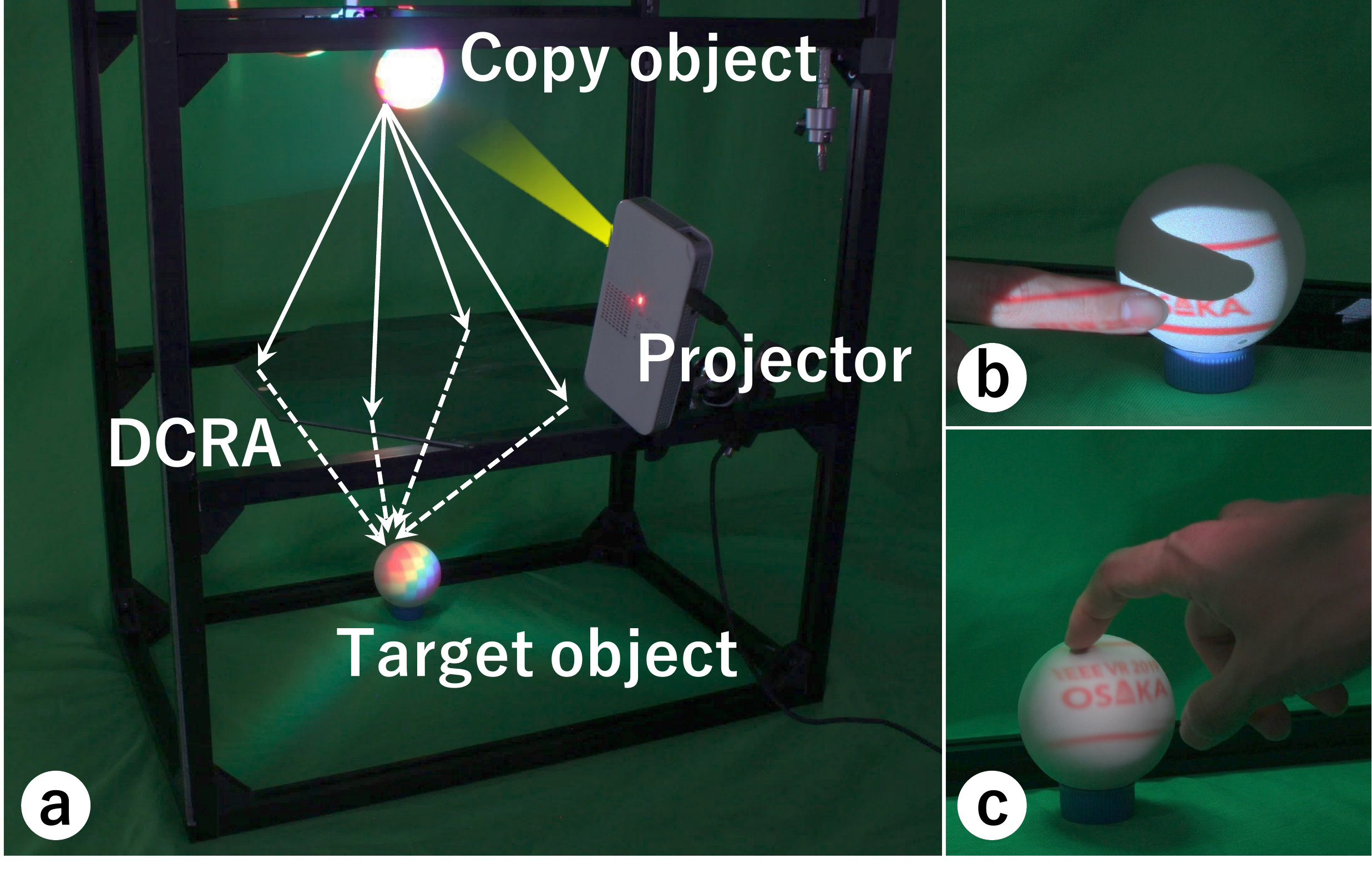}
  \caption{Shadow removal by the proposed large aperture projection system. (a) Experimental setup. (b) A finger occludes projection light in a normal projection mapping system. (c) No shadow was cast on the target object touched by a finger.}
  \label{fig3}
\end{figure}

Shadow removal in projection mapping has been extensively explored~\cite{4270468,Iwai2014,1272728,https://doi.org/10.1111/cgf.13541,Nagase2011,10.1145/3415255.3422888,5544603,990943,10.1145/1056808.1057076,7164338,https://doi.org/10.1111/cgf.13085}. Most of the techniques developed in the past applied multiple projectors to remove shadows. More specifically, suppose a surface area is occluded from a projector, then another projector projects a compensation image onto the occluded surface. This approach works well when an occluder is static. However, due to the delay from the occlusion to the compensation projection, the shadow is not perfectly removed when the occluder moves. To overcome this limitation, we propose an optical solution. A shadow does not occur when the lens aperture of a projector is large enough. Therefore, we build a projection system whose aperture is large~\cite{8798245}. We applied a dihedral corner reflector array (DCRA) to our system~\cite{aska3d,10.1117/12.690574}. DCRA collects light from a point light source to the plane symmetrical point. Therefore, we prepared an object whose shape is the same as the projection target (hereinafter, we call this object a copy object) and placed it at the plane symmetrical position regarding the DCRA (\autoref{fig3}a). Then, the appearance of the copy object is transferred to the target. When we apply a large-format DCRA, then we can consider the aperture of this projection system as very large. In our preliminary experiment, we confirmed that a user’s finger did not cause a hard shadow on the projection target even when the finger touched the object (\autoref{fig3}c). As a reference, when we projected the same texture onto the target object from a standard projector, the shadow occurred when the user’s finger approached the object (\autoref{fig3}b).

\section{Conclusions}

This invited talk introduced our technical solutions to overcome the limitations causing unnatural projection mapping experiences. In particular, we showed a multi-projection approach to reproduce the light field of room lights in a projection mapping application. We also demonstrated our optical approach to solve the shadow problem. I believe that these techniques will expand the application fields of projection mapping that allow users to experience AR without wearing nor holding any devices.

\acknowledgments{
This work was supported by JSPS KAKENHI Grant Numbers JP20H05958 and JST, PRESTO Grant Number JPMJPR19J2, Japan.}

\bibliographystyle{abbrv-doi}

\bibliography{template}

\begin{thebibliography}{10}

\bibitem{aska3d}
{ASKA3D}.
\newblock \url{https://aska3d.com/}.
\newblock Accessed: 2021-12-1.

\bibitem{4270468}
S.~Audet and J.~R. Cooperstock.
\newblock Shadow removal in front projection environments using object
  tracking.
\newblock In {\em 2007 IEEE Conference on Computer Vision and Pattern
  Recognition}, pp. 1--8, 2007.

\bibitem{970539}
D.~Bandyopadhyay, R.~Raskar, and H.~Fuchs.
\newblock Dynamic shader lamps : painting on movable objects.
\newblock In {\em Proceedings IEEE and ACM International Symposium on Augmented
  Reality}, pp. 207--216, 2001.

\bibitem{doi:10.1111/cgf.13128}
A.~H. Bermano, M.~Billeter, D.~Iwai, and A.~Grundhöfer.
\newblock Makeup lamps: Live augmentation of human faces via projection.
\newblock {\em Computer Graphics Forum}, 36(2):311--323, 2017.

\bibitem{1377099}
O.~Bimber, F.~Coriand, A.~Kleppe, E.~Bruns, S.~Zollmann, and T.~Langlotz.
\newblock Superimposing pictorial artwork with projected imagery.
\newblock {\em IEEE MultiMedia}, 12(1):16--26, 2005.

\bibitem{10.5555/1088894}
O.~Bimber and R.~Raskar.
\newblock {\em Spatial Augmented Reality: Merging Real and Virtual Worlds}.
\newblock A. K. Peters, Ltd., 2005.

\bibitem{1544657}
O.~{Bimber}, G.~{Wetzstein}, A.~{Emmerling}, and C.~{Nitschke}.
\newblock Enabling view-dependent stereoscopic projection in real environments.
\newblock In {\em Fourth IEEE and ACM International Symposium on Mixed and
  Augmented Reality (ISMAR'05)}, pp. 14--23, 2005.

\bibitem{CASCINI2020103308}
G.~Cascini, J.~O'Hare, E.~Dekoninck, N.~Becattini, J.-F. Boujut, F.~{Ben
  Guefrache}, I.~Carli, G.~Caruso, L.~Giunta, and F.~Morosi.
\newblock Exploring the use of ar technology for co-creative product and
  packaging design.
\newblock {\em Computers in Industry}, 123:103308, 2020.

\bibitem{10.1145/1166253.1166290}
M.~Flagg and J.~M. Rehg.
\newblock Projector-guided painting.
\newblock In {\em Proceedings of the 19th Annual ACM Symposium on User
  Interface Software and Technology}, UIST '06, p. 235–244. Association for
  Computing Machinery, New York, NY, USA, 2006.

\bibitem{doi:10.1111/cgf.13387}
A.~Grundhöfer and D.~Iwai.
\newblock Recent advances in projection mapping algorithms, hardware and
  applications.
\newblock {\em Computer Graphics Forum}, 37(2):653--675, 2018.

\bibitem{8798245}
K.~Hiratani, D.~Iwai, P.~Punpongsanon, and K.~Sato.
\newblock Shadowless projector: Suppressing shadows in projection mapping with
  micro mirror array plate.
\newblock In {\em 2019 IEEE Conference on Virtual Reality and 3D User
  Interfaces (VR)}, pp. 1309--1310, 2019.

\bibitem{8172039}
D.~{Iwai}, R.~{Matsukage}, S.~{Aoyama}, T.~{Kikukawa}, and K.~{Sato}.
\newblock Geometrically consistent projection-based tabletop sharing for remote
  collaboration.
\newblock {\em IEEE Access}, 6:6293--6302, 2018.

\bibitem{Iwai2014}
D.~Iwai, M.~Nagase, and K.~Sato.
\newblock Shadow removal of projected imagery by occluder shape measurement in
  a multiple overlapping projection system.
\newblock {\em Virtual Reality}, 18(4):245--254, Nov 2014.

\bibitem{Iwai2011}
D.~Iwai and K.~Sato.
\newblock Document search support by making physical documents transparent in
  projection-based mixed reality.
\newblock {\em Virtual Reality}, 15(2):147--160, Jun 2011.

\bibitem{1272728}
C.~Jaynes, S.~Webb, and R.~Steele.
\newblock Camera-based detection and removal of shadows from interactive
  multiprojector displays.
\newblock {\em IEEE Transactions on Visualization and Computer Graphics},
  10(3):290--301, 2004. doi: {{%
10\hspace{.1pt}\discretionary{.}{%
}{.}\hspace{.4pt}1109\discretionary{/}{%
}{/}TVCG\hspace{.1pt}\discretionary{.}{%
}{.}\hspace{.4pt}2004\hspace{.1pt}\discretionary{.}{%
}{.}\hspace{.4pt}1272728}}


\bibitem{https://doi.org/10.1111/cgf.13541}
J.~Kim, H.~Seo, S.~Cha, and J.~Noh.
\newblock Real-time human shadow removal in a front projection system.
\newblock {\em Computer Graphics Forum}, 38(1):443--454, 2019.

\bibitem{8007248}
Y.~Kitajima, D.~Iwai, and K.~Sato.
\newblock Simultaneous projection and positioning of laser projector pixels.
\newblock {\em IEEE Transactions on Visualization and Computer Graphics},
  23(11):2419--2429, 2017.

\bibitem{kijima07}
D.~Kondo, R.~Kijima, and Y.~Takahashi.
\newblock Dynamic anatomical model for medical education using free form
  projection display.
\newblock In {\em Proceedings of the 13th International Conference on Virtual
  Systems and Multimedia (VSMM)}, 2007.

\bibitem{10.1117/12.690574}
S.~Maekawa, K.~Nitta, and O.~Matoba.
\newblock {Transmissive optical imaging device with micromirror array}.
\newblock In B.~Javidi, F.~Okano, and J.-Y. Son, eds., {\em Three-Dimensional
  TV, Video, and Display V}, vol. 6392, pp. 130 -- 137. International Society
  for Optics and Photonics, SPIE, 2006.

\bibitem{6949562}
M.~R. Marner, R.~T. Smith, J.~A. Walsh, and B.~H. Thomas.
\newblock Spatial user interfaces for large-scale projector-based augmented
  reality.
\newblock {\em IEEE Computer Graphics and Applications}, 34(6):74--82, 2014.
  doi: {{%
10\hspace{.1pt}\discretionary{.}{%
}{.}\hspace{.4pt}1109\discretionary{/}{%
}{/}MCG\hspace{.1pt}\discretionary{.}{%
}{.}\hspace{.4pt}2014\hspace{.1pt}\discretionary{.}{%
}{.}\hspace{.4pt}117}}


\bibitem{10.1145/1959826.1959828}
K.~Matsushita, D.~Iwai, and K.~Sato.
\newblock Interactive bookshelf surface for in situ book searching and storing
  support.
\newblock In {\em Proceedings of the 2nd Augmented Human International
  Conference}, 2011.

\bibitem{10.1111:j.1467-8659.2011.02066.x}
C.~Menk, E.~Jundt, and R.~Koch.
\newblock {Visualisation Techniques for Using Spatial Augmented Reality in the
  Design Process of a Car}.
\newblock {\em Computer Graphics Forum}, 2011.

\bibitem{6193074}
M.~R. {Mine}, J.~{van Baar}, A.~{Grundhofer}, D.~{Rose}, and B.~{Yang}.
\newblock Projection-based augmented reality in disney theme parks.
\newblock {\em Computer}, 45(7):32--40, 2012.

\bibitem{Nagase2011}
M.~Nagase, D.~Iwai, and K.~Sato.
\newblock Dynamic defocus and occlusion compensation of projected imagery by
  model-based optimal projector selection in multi-projection environment.
\newblock {\em Virtual Reality}, 15(2):119--132, Jun 2011.

\bibitem{00000658-201806000-00024}
H.~Nishino, E.~Hatano, S.~Seo, T.~Nitta, T.~Saito, M.~Nakamura, K.~Hattori,
  M.~Takatani, H.~Fuji, K.~Taura, and S.~Uemoto.
\newblock Real-time navigation for liver surgery using projection mapping with
  indocyanine green fluorescence: Development of the novel medical imaging
  projection system.
\newblock {\em Annals of Surgery}, 267(6):1134--1140, 2018.

\bibitem{10.1145/3415255.3422888}
T.~Nomoto, W.~Li, H.-L. Peng, and Y.~Watanabe.
\newblock Dynamic projection mapping with networked multi-projectors based on
  pixel-parallel intensity control.
\newblock In {\em SIGGRAPH Asia 2020 Emerging Technologies}. Association for
  Computing Machinery, New York, NY, USA, 2020.

\bibitem{10.1145/2818048.2819965}
T.~Pejsa, J.~Kantor, H.~Benko, E.~Ofek, and A.~Wilson.
\newblock Room2room: Enabling life-size telepresence in a projected augmented
  reality environment.
\newblock In {\em Proceedings of the 19th ACM Conference on Computer-Supported
  Cooperative Work and Social Computing}, CSCW '16, p. 1716–1725, 2016.

\bibitem{10.1145/1015706.1015738}
R.~Raskar, P.~Beardsley, J.~van Baar, Y.~Wang, P.~Dietz, J.~Lee, D.~Leigh, and
  T.~Willwacher.
\newblock Rfig lamps: Interacting with a self-describing world via photosensing
  wireless tags and projectors.
\newblock {\em ACM Trans. Graph.}, 23(3):406–415, aug 2004.

\bibitem{10.1145/280814.280861}
R.~Raskar, G.~Welch, M.~Cutts, A.~Lake, L.~Stesin, and H.~Fuchs.
\newblock The office of the future: A unified approach to image-based modeling
  and spatially immersive displays.
\newblock In {\em Proceedings of the 25th Annual Conference on Computer
  Graphics and Interactive Techniques}, SIGGRAPH '98, p. 179–188, 1998.

\bibitem{10.1145/2366145.2366176}
A.~Rivers, A.~Adams, and F.~Durand.
\newblock Sculpting by numbers.
\newblock {\em ACM Trans. Graph.}, 31(6), nov 2012.

\bibitem{SCHMIDT20191}
S.~Schmidt, G.~Bruder, and F.~Steinicke.
\newblock Effects of virtual agent and object representation on experiencing
  exhibited artifacts.
\newblock {\em Computers \& Graphics}, 83:1 -- 10, 2019.

\bibitem{8007312}
C.~{Siegl}, V.~{Lange}, M.~{Stamminger}, F.~{Bauer}, and J.~{Thies}.
\newblock Faceforge: Markerless non-rigid face multi-projection mapping.
\newblock {\em IEEE Transactions on Visualization and Computer Graphics},
  23(11):2440--2446, 2017.

\bibitem{5544603}
Y.~Sugaya, I.~Miyagawa, and H.~Koike.
\newblock Contrasting shadow for occluder light suppression from one-shot
  image.
\newblock In {\em 2010 IEEE Computer Society Conference on Computer Vision and
  Pattern Recognition - Workshops}, pp. 96--103, 2010.

\bibitem{990943}
R.~Sukthankar, T.-J. Cham, and G.~Sukthankar.
\newblock Dynamic shadow elimination for multi-projector displays.
\newblock In {\em Proceedings of the 2001 IEEE Computer Society Conference on
  Computer Vision and Pattern Recognition. CVPR 2001}, vol.~2, pp. II--II,
  2001.

\bibitem{10.1145/1056808.1057076}
J.~Summet, G.~D. Abowd, G.~M. Corso, and J.~M. Rehg.
\newblock Virtual rear projection: Do shadows matter?
\newblock In {\em CHI '05 Extended Abstracts on Human Factors in Computing
  Systems}, CHI EA '05, p. 1997–2000. Association for Computing Machinery,
  New York, NY, USA, 2005.

\bibitem{8797923}
T.~Takezawa, D.~Iwai, K.~Sato, T.~Hara, Y.~Takeda, and K.~Murase.
\newblock Material surface reproduction and perceptual deformation with
  projection mapping for car interior design.
\newblock In {\em 2019 IEEE Conference on Virtual Reality and 3D User
  Interfaces (VR)}, pp. 251--258, 2019.

\bibitem{7164338}
J.~Tsukamoto, D.~Iwai, and K.~Kashima.
\newblock Radiometric compensation for cooperative distributed multi-projection
  system through 2-dof distributed control.
\newblock {\em IEEE Transactions on Visualization and Computer Graphics},
  21(11):1221--1229, 2015.

\bibitem{https://doi.org/10.1111/cgf.13085}
J.~Tsukamoto, D.~Iwai, and K.~Kashima.
\newblock Distributed optimization framework for shadow removal in
  multi-projection systems.
\newblock {\em Computer Graphics Forum}, 36(8):369--379, 2017.

\bibitem{10.1145/302979.303114}
J.~Underkoffler and H.~Ishii.
\newblock Urp: A luminous-tangible workbench for urban planning and design.
\newblock In {\em Proceedings of the SIGCHI Conference on Human Factors in
  Computing Systems}, p. 386–393. Association for Computing Machinery, New
  York, NY, USA, 1999.

\end{thebibliography}
\end{document}